\documentstyle[sprocl]{article}

\input{psfig}

\begin{document}

\title{WHY DO SOLAR NEUTRINO EXPERIMENTS BELOW \hbox{1
MEV?}\footnote{To be published in the proceedings of the SEcond International
Workshop on Low Energy Solar Neutrinos, University of Tokyo, Tokyo,
Japan, December 4 and 5, 2000 (World Scientific).}}

\author{J. N. BAHCALL}

\address{Institute for Advanced Study, Princeton, NJ 08540, USA\\
E-mail: jnb@sns.ias.edu}

\maketitle\abstracts{ I discuss why we need solar neutrino experiments
below 1 MeV. I also express my prejudices about the desired number and types
of such experiments, emphasizing the importance of $p$-$p$ solar neutrino
experiments.}

The great challenge of solar neutrino research is to make accurate
measurements of neutrinos with energies less than $1$ MeV. We need to
develop experiments that will measure the the total flux, the flavor
content, and the time dependence of the $^7$Be neutrinos (energy of
$0.86$ MeV), and the total flux, flavor content, energy spectrum, and
time dependence of the fundamental $p$-$p$ neutrinos ($< 0.43$ MeV).

More than $98$\% of the calculated standard model solar neutrino flux
lies below $1$ MeV. The rare $^8$B neutrino flux is the only solar
neutrino source for which measurements of the energy have been made,
but $^8$B neutrinos constitute a fraction of less than $10^{-4}$ of
the total solar neutrino flux.

The $p$-$p$  neutrinos are overwhelmingly the most abundant source of
solar neutrinos, carrying about $91$\% of the total flux according to
the standard solar model. The $^7$Be neutrinos constitute about $7$\%
of the total standard model flux.

I want to express first my own views about what we should and should not
emphasize in developing new experiments and then say a little bit
about specific experiments.

Each of the measurable quantities for low energy solar neutrinos is
important and can be used to constrain models of the neutrino and of
the sun. In my view, too much emphasis has been placed in the past on
trying to devise experiments that can do everything. I think we should
be happy if a low energy solar neutrino experiment can measure any of
the desired physical quantities accurately. For example, an experiment
that is sensitive to time dependences need not necessarily measure a
flux accurately. If an experiment measures a charged current rate, it
does not need to provide detailed spectral information. We have to
learn how to crawl before we try to run.

We should aim at ultimately developing experiments with high
statistical significance in order to refine the tests of solar models
and neutrino oscillations. But, the first experiments do not have to
have high counting rates, especially if they are modular and can
demonstrate proof-of-principle.

The interaction cross sections must be known accurately, to a
$1\sigma$ accuracy of $\sim \pm 5\%$ or better, if we are to have a
measurement that is good to $\pm 10\%$. I think a $1\sigma$
measurement of the total rate,  for $^7$Be and for $p$-$p$ neutrinos, that
is at least as accurate as $\pm 10\%$ is necessary in order to make
real progress. There is no reason to believe that we can rely on (p,n)
measurements or nuclear model calculations to provide a determination
of the absolute cross section to this accuracy. Instead, we must
either make use of the related beta-decay process when available or
carry out precise measurements with intense radioactive sources.

Solar neutrino experiments are all difficult and all take a very long
time to carry out. It is tempting to say that a given part of
parameter space is covered by a particular experiment and so we must
design an experiment that tests an entirely different part of
parameter space. I think this type of reasoning is dangerous, because
the history of science shows that experimental results are
misinterpreted or are misleading much more often than one would expect
from the quoted errors. Moreover, the claim that two different
experimental techniques measure the same quantity often rests upon a
theoretical assumption, a theoretical model that itself requires
testing.

We must have redundancy. We must have different ways of
measuring the same quantities. The implications of the experimental
results, for physics and for astronomy, are too important to depend
upon single experiments.

A number of promising possibilities were discussed at the LowNu2
workshop. These include the BOREXINO observatory, which can detect
$\nu-e$ scattering and is so far the only approved solar neutrino
experiment that is both being built at full scale and that can measure
neutrino energies less than $1$ MeV. Other very promising experiments
that were described at this workshop include CLEAN, GENIUS, HERON,
KamLAND, LENS, MOON, and XMASS. After the workshop,
Raju Raghavan$\,$\cite{raghavan} succeeded in demonstrating that one can build
a stable In liquid scintillator that could potentially be used for a
very low threshold $p-p$ solar neutrino detector (if one can overcome
by coincidence and modular techniques the unfavorable raw signal to
noise ratio of $10^{-11}$).

We want to test and to understand neutrino oscillations with high
precision using solar neutrino sources. 

Magic things can be done with neutrino lines$\,\,$\cite{jnbplamen}, like the
$0.86{\rm ~ MeV ~} ^7$Be line. To make the magic work, one has to
measure the neutrino-electron scattering rate (as will be done for the
$^7$Be line with the BOREXINO experiment), and also the CC
(neutrino-absorption) rate with the same line (no approved
experiment). Assuming there are no sterile neutrinos, one can then use
the two measurements to determine uniquely the survival probability at
a particular energy and the total neutrino flux. One can test for the
existence of sterile neutrinos by measuring~\cite{jnbplamen} the
neutrino-electron scattering rate and the CC rate for both the $0.86$
MeV and the $0.34$ MeV $^7$Be neutrino lines, but this is a tough
job.

 The time dependences, seasonal and day-night, of the observed event
rates of the $^7$Be neutrino lines will be valuable diagnostic tests
of neutrino oscillation scenarios.

I believe that we have calculated the flux of $p$-$p$ neutrinos
produced in the sun to an accuracy of $\pm 1$\%. This belief should be
tested experimentally. Unfortunately, we do not yet have a direct
measurement of this flux. The gallium experiments, which have played
an enormously important role in understanding what is happening to
solar neutrinos, nevertheless only tell us the rate of capture
of all neutrinos with energies above $0.23$ MeV.

The most urgent need for solar neutrino research is to develop
practical experiments to measure directly the $p$-$p$ neutrino flux,
hopefully both charged current and neutrino-electron scattering, the
energy spectrum, and the time dependences.  An experiment, or a
combination of different experiments, that measures the total flux of
$p$-$p$ neutrinos can be used to test the precise and fundamental
standard solar model prediction of the \hbox{$p$-$p$} neutrino flux.

\begin{figure}[!t]
\centerline{\psfig{figure=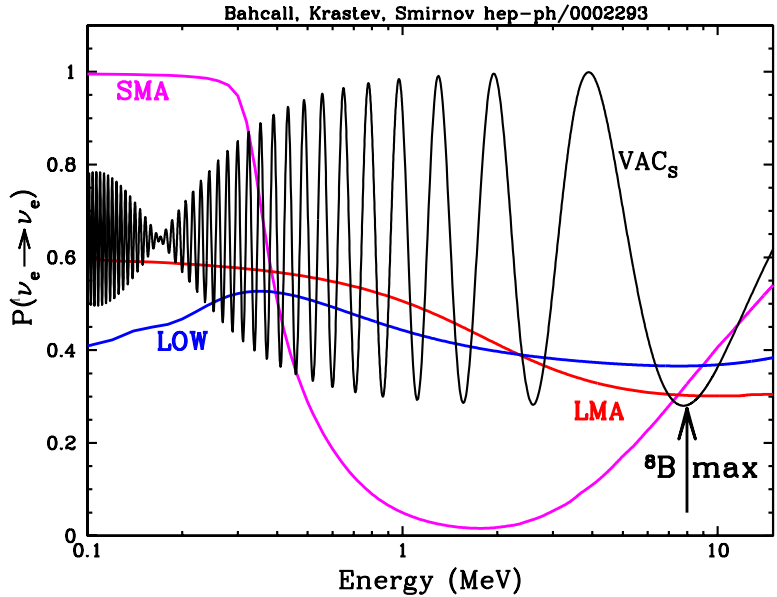,width=4in,angle=270}}
\caption[]{Survival probabilities for MSW solutions.
The figure presents the yearly-averaged 
survival probabilities for an electron neutrino that is  created
in the sun to remain an electron neutrino 
 upon arrival at the Super-Kamiokande
detector. 
\label{fig:survival}}
\end{figure}

Figure~\ref{fig:survival} shows the calculated neutrino survival
probability as a function of energy for three global best-fit MSW
oscillation solutions.  You can see directly from this figure why we
need accurate measurements for the $p$-$p$ and $^7$Be neutrinos.  The
currently favored solutions exhibit their most characteristic and
strongly energy dependent features below $1$ MeV.  Naturally, all of
the solutions give similar predictions in the energy region, $\sim 7$
MeV, where the Kamiokande, Super-Kamiokande, and SNO data are best.  The
survival probability shows a strong change with energy below $1$ MeV
for all the solutions, whereas in the region above $5$ MeV (accessible
to Super-Kamiokande and to SNO) the energy dependence of the survival
probability is at best modest.

Measurements of both the CC and the neutrino-electron scattering rate
of either the $^7$Be or the $p$-$p$ neutrinos will be extremely
important. When combined, they can determine the total neutrino flux
and therefore allow a direct comparison with solar model
predictions. The same thing could be achieved by a neutral current
measurement, although that may be more difficult to obtain in practice.
 
In the more distant future, we will want to measure the average energy
and shape of the $^7$Be neutrino line with a precision better than
$0.3$ keV in order to obtain a direct determination of the central
temperature of the sun. The standard solar model predicts that the
average energy of the $^7$Be neutrinos emitted from the sun exceeds by
$1.3$ keV the laboratory energy of the (higher energy) $^7$Be
line. This energy  shift is due to the high temperature of the plasma in the
region in which the $^7$Be line is produced.

The $p$-$p$ neutrinos are the gold ring of solar neutrino physics and
astronomy.  Their measurement will constitute a simultaneous and
critical test of stellar evolution theory and of neutrino oscillation
solutions. 

No matter what we learn from experiments at higher neutrino
energies, from the wonderful experiments of SNO and SuperKamiokande,
we will still desperately want to measure the $p$-$p$ neutrinos. The $p$-$p$
neutrinos are a fundamental product of the solar energy generation
process who flux is precisely predicted but not yet measured separately.
The $p$-$p$ neutrinos represent the dominant mode of neutrino emission
from the sun, with a flux that is $10^4$ times larger than the flux of
the rare $^8$B neutrinos measured by SNO and
SuperKamiokande. Therefore, measurements of the $p$-$p$ neutrinos will
severely test theoretical ideas regarding both the interior of the sun
and the nature of neturinos that are inferred from measurements of the
less abundant, higher energy neutrinos.

\section*{Acknowledgments}
I am grateful to Professor Yoichiro Suzuki for his leadership in
organizing this workshop and for his insightful guidance in understanding
the nature of low energy neutrino experiments.
I acknowledge support from NSF grant \#PHY0070928.

\end{document}